\documentstyle[prd,preprint,tighten,aps,eqsecnum,floats,%
amssymb,amsfonts,newlfont,graphicx]{revtex}
\begin{document}
\draft
\preprint{
\begin{tabular}{r}
DFTT 54/99\\
hep-ph/9910336
\end{tabular}
}
\title{Four-Neutrino Oscillations}
\author{Carlo Giunti}
\address{INFN, Sezione di Torino, and Dipartimento di Fisica Teorica,
Universit\`a di Torino,\\
Via P. Giuria 1, I--10125 Torino, Italy}
\maketitle
\begin{abstract}
It is shown that at least four massive neutrinos are
needed in order to accommodate the evidences in favor of neutrino oscillations
found in solar and atmospheric neutrino experiments
and in the LSND experiment.
Among all four-neutrino schemes,
only two,
with a mass spectrum composed of
two pairs of neutrinos with close masses 
separated by the ``LSND gap'' of the order of 1 eV,
are compatible with the results
of all neutrino oscillation experiments.
In these two schemes the probability
of
$
\stackrel{\makebox[0pt][l]
{$\hskip-3pt\scriptscriptstyle(-)$}}{\nu_{e}}
$
transitions into other states,
the probability of
$
\stackrel{\makebox[0pt][l]
{$\hskip-3pt\scriptscriptstyle(-)$}}{\nu_{\mu}}
\to\stackrel{\makebox[0pt][l]
{$\hskip-3pt\scriptscriptstyle(-)$}}{\nu_{e}}
$
transitions
and
the size of CP violation effects in
$\nu_\mu\leftrightarrows\nu_e$
and
$\bar\nu_\mu\leftrightarrows\bar\nu_e$
transitions
are suppressed
in long-baseline experiments.
\end{abstract}

\pacs{Talk presented at
the ICFA/ECFA Workshop
\textit{Neutrino Factories based on Muon Storage Rings},
$\nu$-Fact'99, Lyon, France, 5-9 July 1999.}

\section{Introduction}
\label{Introduction}

The existence of neutrino masses and mixing
is today one of the hottest topics in high-energy physics
and
is considered as one of the best ways to obtain indications
on the physics
beyond the Standard Model.
If neutrinos are massive and mixed,
the left-handed components
$\nu_{{\alpha}L}$
($\alpha=e,\mu,\tau,\ldots$)
of the flavor neutrino fields
are superpositions of
the left-handed components
$\nu_{kL}$
($k=1,\ldots,N$)
of neutrino fields with definite mass
$m_k$,
$ \displaystyle
\nu_{{\alpha}L}
=
\sum_{k=1}^{N}
U_{{\alpha}k}
\,
\nu_{kL}
$,
where $U$
is a $N{\times}N$ unitary mixing matrix.
In this case neutrino oscillations occur.
From the measurement of the invisible decay width of the $Z$-boson
it is known that the number of light active
neutrino flavors is three,
corresponding to $\nu_e$, $\nu_\mu$ and $\nu_\tau$.
This implies that
the number $N$ of massive neutrinos is bigger or equal to three.
If $N>3$, in the flavor basis there are $N_s=N-3$
sterile neutrinos,
$\nu_{s_1}$, \ldots, $\nu_{s_{N_s}}$.
In this case the flavor index $\alpha$
takes the values
$e,\mu,\tau,s_1,\ldots,s_{N_s}$.

Evidences in favor of neutrino oscillations have been found
in solar neutrino experiments \cite{exp-sun},
in atmospheric neutrino experiments \cite{exp-atm}
and in the LSND accelerator experiment \cite{LSND-Moriond-99}.
The observed disappearance of atmospheric
$
\stackrel{\makebox[0pt][l]
{$\hskip-3pt\scriptscriptstyle(-)$}}{\nu_{\mu}}
$'s can be explained by
$
\stackrel{\makebox[0pt][l]
{$\hskip-3pt\scriptscriptstyle(-)$}}{\nu_{\mu}}
\to\stackrel{\makebox[0pt][l]
{$\hskip-3pt\scriptscriptstyle(-)$}}{\nu_{\tau}}
$
and/or
$
\stackrel{\makebox[0pt][l]
{$\hskip-3pt\scriptscriptstyle(-)$}}{\nu_{\mu}}
\to\stackrel{\makebox[0pt][l]
{$\hskip-3pt\scriptscriptstyle(-)$}}{\nu_{s}}
$
transitions,
the observed disappearance of solar $\nu_e$'s can be explained by
$\nu_e\to\nu_\mu$
and/or
$\nu_e\to\nu_\tau$
and/or
$\nu_e\to\nu_s$
transitions,
and
$\bar\nu_\mu\to\bar\nu_e$ and $\nu_\mu\to\nu_e$
transitions
have been observed in the LSND experiment.

\section{The necessity of at least three independent
$\Delta{\lowercase{\mathbf{m}}}^2$'s}
\label{necessity}

The three evidences in favor of neutrino oscillations
found in solar and atmospheric neutrino experiments
and in the accelerator LSND experiment
imply the existence of at least three independent
neutrino mass-squared differences.
This can be seen by considering
the general expression for the probability of
$\nu_\alpha\to\nu_\beta$
transitions in vacuum,
that can be written as
(see \cite{BGG-review-98})
\begin{equation}
P_{\nu_\alpha\to\nu_\beta}
=
\left|
\sum_{k=1}^{N}
U_{{\alpha}k}^* \,
U_{{\beta}k} \,
\exp\left( - i \, \frac{ \Delta{m}^2_{kj} \, L }{ 2 \, E } \right)
\right|^2
\,,
\label{Posc}
\end{equation}
where
$ \Delta{m}^2_{kj} \equiv m_k^2-m_j^2 $,
$j$ is any of the mass-eigenstate indices,
$L$ is the distance between the neutrino source and detector
and $E$ is the neutrino energy.
The range of $L/E$ characteristic of each type of experiment is different:
$ L / E \sim 10^{11} - 10^{12} \, \mathrm{eV}^{-2} $
for solar neutrino experiments,
$ L / E \sim 10^{2} - 10^{3} \, \mathrm{eV}^{-2} $
for atmospheric neutrino experiments
and
$ L / E \sim 1 \, \mathrm{eV}^{-2} $
for the LSND experiment.
From Eq.~(\ref{Posc}) it is clear that neutrino oscillations
are observable in an experiment only if there is at least one
mass-squared difference
$\Delta{m}^2_{kj}$ such that
$ \Delta{m}^2_{kj} L / 2 E \gtrsim 0.1 $
(the precise lower bound depends on the sensitivity of the experiment)
in a significant part of the energy and source-detector distance
intervals of the experiment
(if this condition is not satisfied,
$
P_{\nu_\alpha\to\nu_\beta}
\simeq
\left|
\sum_k
U_{{\alpha}k}^* \,
U_{{\beta}k}
\right|^2
=
\delta_{\alpha\beta}
$).
Since the range of $L/E$ probed by the LSND experiment is the smaller one,
a large mass-squared difference is needed for LSND oscillations,
$
\Delta{m}^2_{\mathrm{LSND}} \gtrsim 10^{-1} \, \mathrm{eV}^2
$.
The 99\% CL maximum likelihood analysis of the LSND data
in terms of two-neutrino oscillations
gives \cite{LSND-Moriond-99}
\begin{equation}
0.20 \, \mathrm{eV}^2
\lesssim
\Delta{m}^2_{\mathrm{LSND}}
\lesssim
2.0 \, \mathrm{eV}^2
\,.
\label{LSND-range}
\end{equation}

Furthermore,
from Eq.~(\ref{Posc}) it is clear that 
a dependence of the oscillation probability
from the neutrino energy $E$
and the source-detector distance $L$
is observable only if there is at least one
mass-squared difference
$\Delta{m}^2_{kj}$ such that
$ \Delta{m}^2_{kj} L / 2 E \sim 1 $.
Indeed,
the exponentials of all the phases
$ \Delta{m}^2_{kj} L / 2 E \ll 1 $
are equal to one
and the contributions of
all the phases
$ \Delta{m}^2_{kj} L / 2 E \gg 1 $
are washed out by the average over the energy and source-detector
ranges characteristic of the experiment.
Since
a variation of the oscillation probability as a function of neutrino energy
has been observed both in solar and atmospheric neutrino experiments
and the ranges of $L/E$ characteristic
of these two types of experiments are different from each other
and different from the LSND range,
two more mass-squared differences with different scales are needed:
\begin{equation}
\Delta{m}^2_{\mathrm{sun}} \sim 10^{-12} - 10^{-11} \, \mathrm{eV}^2
\quad
\mbox{(VO)}
\,,
\qquad
\Delta{m}^2_{\mathrm{atm}} \sim 10^{-3} - 10^{-2} \, \mathrm{eV}^2
\,.
\label{dm2-sun-atm}
\end{equation}
The condition (\ref{dm2-sun-atm}) for the solar mass-squared difference
$\Delta{m}^2_{\mathrm{sun}}$
has been obtained under the assumption of vacuum oscillations (VO). 
If the disappearance of solar $\nu_e$'s is due to the MSW effect
(see \cite{BGG-review-98}),
the condition
\begin{equation}
\Delta{m}^2_{\mathrm{sun}} \lesssim 10^{-4} \, \mathrm{eV}^2
\qquad
\mbox{(MSW)}
\label{dm2-sun-MSW}
\end{equation}
must be fulfilled
in order to have a resonance in the interior of the sun.
Hence,
in the MSW case
$\Delta{m}^2_{\mathrm{sun}}$
must be at least one order of magnitude smaller than
$\Delta{m}^2_{\mathrm{atm}}$.

It is possible to ask if three different scales
of neutrino mass-squared differences are needed even
if the results of the Homestake solar neutrino experiment
is neglected, allowing an energy-independent suppression of the solar $\nu_e$ flux.
The answer is that still the data cannot be fitted with only
two neutrino mass-squared differences
because an energy-independent suppression of the solar $\nu_e$ flux
requires large $\nu_e\to\nu_\mu$ or $\nu_e\to\nu_\tau$
transitions generated by $\Delta{m}^2_{\mathrm{atm}}$
or $\Delta{m}^2_{\mathrm{LSND}}$.
These transitions
are forbidden by the results of the Bugey \cite{Bugey-95}
and CHOOZ \cite{CHOOZ}
reactor $\bar\nu_e$ disappearance experiments
and by the non-observation of an up-down asymmetry
of $e$-like events in the Super-Kamiokande
atmospheric neutrino experiment \cite{SK-atm}.

\section{Four-neutrino schemes}
\label{Four-neutrino schemes}

The existence of three different scales of $\Delta{m}^2$
imply that at least four light massive neutrinos must exist in nature.
Here we consider the schemes with four light and mixed neutrinos,
which constitute the minimal possibility
that allows to accommodate the results
of all neutrino oscillation experiments.
In this case,
in the flavor basis the three active neutrinos $\nu_e$, $\nu_\mu$, $\nu_\tau$
are accompanied by a sterile neutrino $\nu_s$.

\begin{figure}[t]
\begin{center}
\includegraphics[bb=13 668 522 827,width=0.99\linewidth]{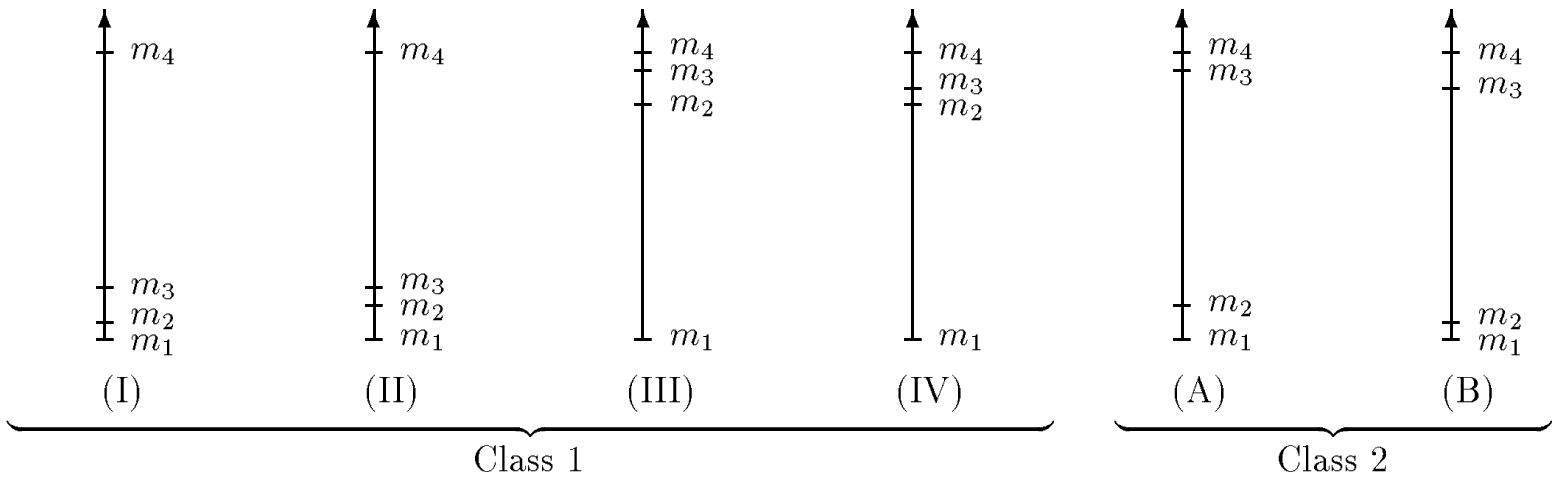}
\refstepcounter{figure}
\label{4spectra}
\small
Figure \ref{4spectra}
\end{center}
\end{figure}

The six types of four-neutrino mass spectra
with
three different scales of $\Delta{m}^2$
that can accommodate
the hierarchy
$
\Delta{m}^2_{\mathrm{sun}}
\ll
\Delta{m}^2_{\mathrm{atm}}
\ll
\Delta{m}^2_{\mathrm{LSND}}
$
are shown qualitatively in Fig.~\ref{4spectra}.
In all these mass spectra there are two groups
of close masses separated by the ``LSND gap'' of the order of 1 eV.
In each scheme the smallest mass-squared
difference corresponds to
$\Delta{m}^2_{\mathrm{sun}}$
($\Delta{m}^2_{21}$ in schemes I and B,
$\Delta{m}^2_{32}$ in schemes II and IV,
$\Delta{m}^2_{43}$ in schemes III and A),
the intermediate one to
$\Delta{m}^2_{\mathrm{atm}}$
($\Delta{m}^2_{31}$ in schemes I and II,
$\Delta{m}^2_{42}$ in schemes III and IV,
$\Delta{m}^2_{21}$ in scheme A,
$\Delta{m}^2_{43}$ in scheme B)
and the largest mass squared difference
$ \Delta{m}^2_{41} = \Delta{m}^2_{\mathrm{LSND}} $
is relevant for the oscillations observed in the LSND experiment.
The six schemes are divided into four schemes of class 1 (I--IV)
in which there is a group of three masses separated from an isolated mass
by the LSND gap,
and two schemes of class 2 (A, B)
in which there are two couples of close masses separated by the LSND gap.

It has been show that
the schemes of class 1 are disfavored by the data
if also the negative
results of short-baseline $\bar\nu_e$ and $\nu_\mu$ disappearance
experiments are taken into account
\cite{BGG-AB-96,Barger-variations-98,BGGS-AB-99,Giunti-70-99}.
This is basically due to the fact
that the non-observation of neutrino oscillations
due to $\Delta{m}^2_{41}$
in short-baseline disappearance experiments
imply that,
in each scheme in Fig.~\ref{4spectra},
$\nu_e$ and $\nu_\mu$
are mainly superpositions of one of the two groups
of mass eigenstates separated by the LSND gap.
Hence,
in the schemes of class 1
$\nu_e$ and $\nu_\mu$
almost coincide with superpositions of the three grouped mass eigenstates
or with the isolated mass eigenstate.
Moreover,
only the possibility of both
$\nu_e$ and $\nu_\mu$
mainly superpositions of the three grouped mass eigenstates
allows to explain the results of solar and atmospheric neutrino experiments
with neutrino oscillations.
This is because
disappearance of solar $\nu_e$'s
and atmospheric $\nu_\mu$'s
is possible only if
$\nu_e$ and $\nu_\mu$
have large mixing with the mass eigenstates
whose mass-squared differences
give
$\Delta{m}^2_{\mathrm{sun}}$
and
$\Delta{m}^2_{\mathrm{atm}}$.
In all schemes of class 1
$\Delta{m}^2_{\mathrm{sun}}$
and
$\Delta{m}^2_{\mathrm{atm}}$
are mass-squared differences between
two of the three grouped mass eigenstates neutrinos.
However,
if both $\nu_e$ and $\nu_\mu$ are
mainly superpositions of the three grouped mass eigenstates,
short-baseline $\nu_\mu\to\nu_e$ oscillations due to
$\Delta{m}^2_{41}$
are strongly suppressed
and one can calculate that the allowed transition probability
is smaller than that observed in the LSND experiment
\cite{BGG-AB-96,BGGS-AB-99}.
Hence,
we conclude that the schemes of class 1
are disfavored by neutrino oscillation data.

\begin{figure}[t]
\begin{tabular*}{\linewidth}{@{\extracolsep{\fill}}cc}
\begin{minipage}{0.47\linewidth}
\begin{center}
\includegraphics[bb=60 326 510 741,width=0.99\linewidth]{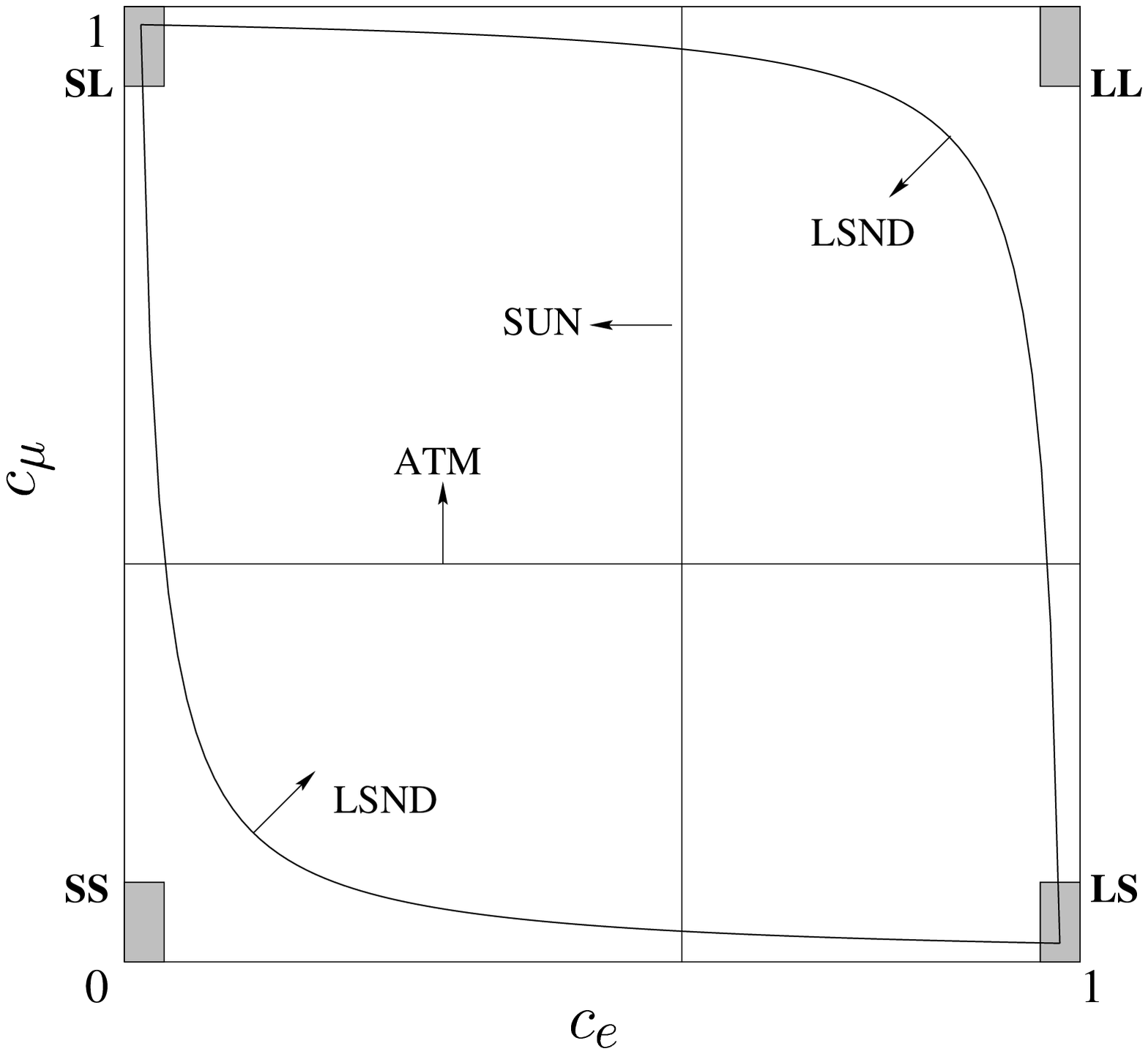}
\refstepcounter{figure}
\label{4dis1}
\small
Figure \ref{4dis1}
\end{center}
\end{minipage}
&
\begin{minipage}{0.47\linewidth}
\begin{center}
\includegraphics[bb=60 326 512 751,width=0.99\linewidth]{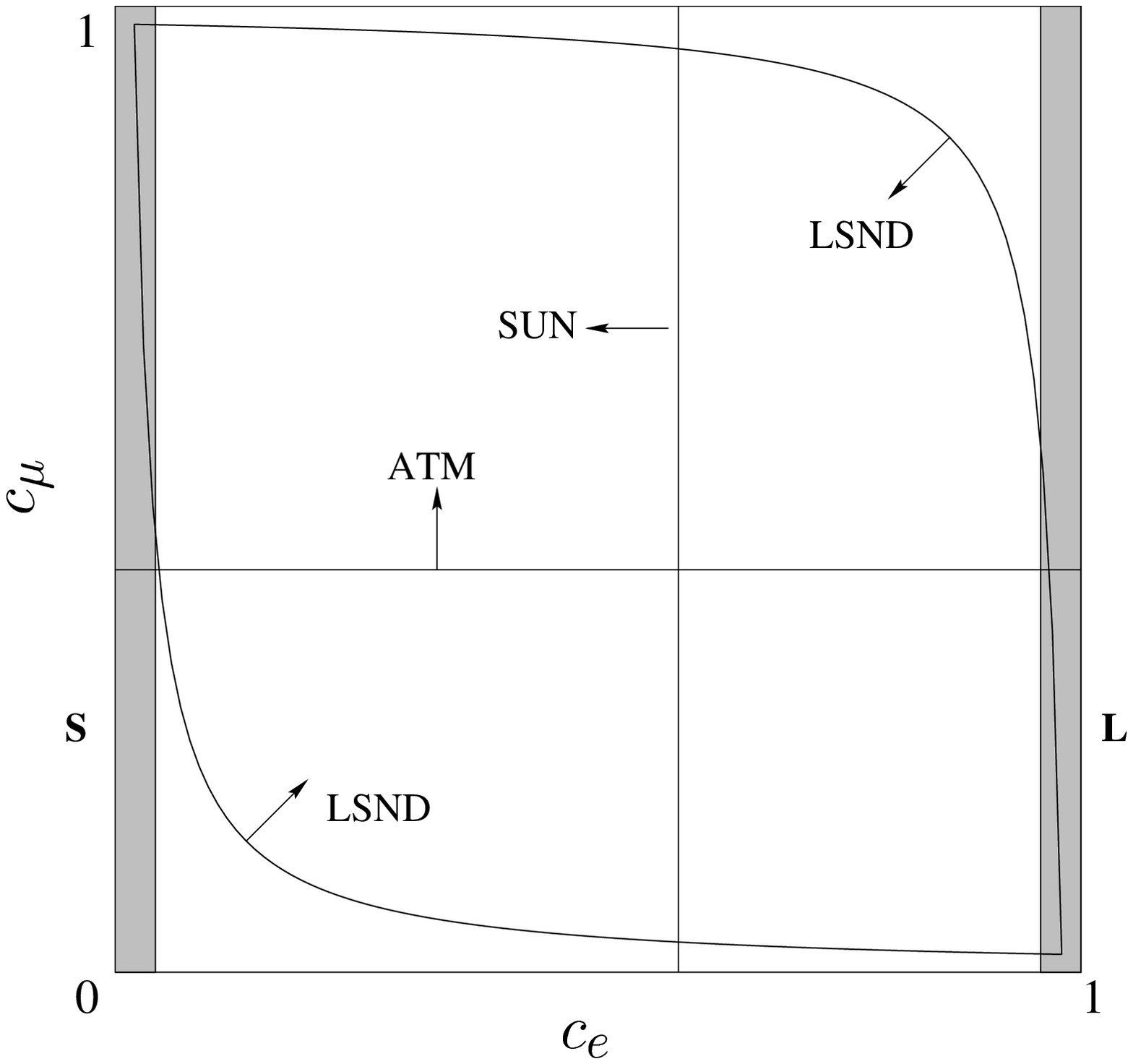}
\refstepcounter{figure}
\label{4dis2}
\small
Figure \ref{4dis2}
\end{center}
\end{minipage}
\end{tabular*}
\end{figure}

The two four-neutrino schemes of class 2
are compatible with the results of all neutrino oscillation experiments
if
the mixing of $\nu_e$ with the two mass eigenstates responsible
for the oscillations of solar neutrinos
($\nu_3$ and $\nu_4$ in scheme A
and
$\nu_1$ and $\nu_2$ in scheme B)
is large
and
the mixing of $\nu_\mu$ with the two mass eigenstates responsible
for the oscillations of atmospheric neutrinos
($\nu_1$ and $\nu_2$ in scheme A
and
$\nu_3$ and $\nu_4$ in scheme B)
is large
\cite{BGKP-96,BGG-AB-96,Barger-variations-98,BGGS-AB-99}.
This is illustrated qualitatively in Figs. \ref{4dis1} and \ref{4dis2},
as we are going to explain.

Let us define the quantities $c_\alpha$,
with $\alpha=e,\mu,\tau,s$,
in the schemes A and B as
\begin{equation}
c_\alpha^{\mathrm{(A)}}
\equiv
\sum_{k=1,2}
|U_{{\alpha}k}|^2
\,,
\qquad
c_\alpha^{\mathrm{(B)}}
\equiv
\sum_{k=3,4}
|U_{{\alpha}k}|^2
\,.
\label{04}
\end{equation}
Physically $c_\alpha$ quantify the mixing of the flavor neutrino $\nu_\alpha$
with the two massive neutrinos whose $\Delta{m}^2$
is relevant for the oscillations of atmospheric neutrinos
($\nu_1$, $\nu_2$ in scheme A
and
$\nu_3$, $\nu_4$ in scheme B).
The negative results of short-baseline
disappearance experiments
imply that \cite{BGG-AB-96}
\begin{equation}
c_\alpha \leq a^0_\alpha
\qquad \mbox{or} \qquad
c_\alpha \geq 1-a^0_\alpha
\qquad
(\alpha=e,\mu)
\,.
\label{c-small-large}
\end{equation}
The quantities
$a^0_e$ and $a^0_\mu$,
that depend on
$\Delta m^2_{41}=\Delta{m}^2_{\mathrm{LSND}}$,
are obtained, respectively,
from the exclusion plots of short-baseline
$\bar\nu_e$ and $\nu_\mu$ experiments
(see \cite{BGG-review-98}).
From the exclusion curves of the Bugey reactor $\bar\nu_e$
disappearance experiment \cite{Bugey-95}
and of the CDHS and CCFR accelerator $\nu_\mu$ disappearance experiments
\cite{CDHS-CCFR-84}
it follows that
$ a_e^0 \lesssim 3 \times 10^{-2} $
for
$\Delta m^2_{41}$
in the LSND range (\ref{LSND-range})
and
$ a_\mu^0 \lesssim 0.2 $
for
$\Delta m^2_{41} \gtrsim 0.4 \, \mathrm{eV}^2$.

The shadowed areas in Figs. \ref{4dis1} and \ref{4dis2}
illustrate qualitatively
the regions in the
$c_e$--$c_\mu$ plane
allowed by the negative results of short-baseline
$\bar\nu_e$ and $\nu_\mu$
disappearance experiments
for a fixed value of
$\Delta m^2_{41}$.
Figure~\ref{4dis1} is valid for
$\Delta m^2_{41} \gtrsim 0.3 \, \mathrm{eV}^2$
and shows that there are four regions
allowed by the results of short-baseline disappearance experiments:
region SS with small $c_e$ and $c_\mu$,
region LS with large $c_e$ and small $c_\mu$,
region SL with small $c_e$ and large $c_\mu$
and
region LL with large $c_e$ and $c_\mu$.
The quantities $c_e$ and $c_\mu$
can be both large,
because the unitarity of the mixing matrix
imply that
$c_\alpha + c_\beta \leq 2$
and
$0 \leq c_\alpha \leq 1$
for $\alpha,\beta=e,\mu,\tau,s$.
Figure~\ref{4dis2} is valid for
$\Delta m^2_{41} \lesssim 0.3 \, \mathrm{eV}^2$,
where there is no constraint on the value of $c_\mu$
from the results of short-baseline
$\nu_\mu$
disappearance experiments.
It shows that there are two regions
allowed by the results of short-baseline $\bar\nu_e$
disappearance experiments:
region S with small $c_e$
and
region L with large $c_e$.

Let us take now into account the results of solar neutrino experiments.
Large values of $c_e$ are incompatible with solar neutrino oscillations
because in this case $\nu_e$ has large mixing with
the two massive neutrinos responsible for atmospheric neutrino oscillations
and, through the unitarity of the mixing matrix,
small mixing with
the two massive neutrinos responsible for solar neutrino oscillations.
Indeed,
in the schemes of class 2
the survival probability
$P^{\mathrm{sun}}_{\nu_e\to\nu_e}$
of solar $\nu_e$'s is bounded by
$
P^{\mathrm{sun}}_{\nu_e\to\nu_e}
\geq
c_e^2 / 2
$,
and its possible variation
$\Delta P^{\mathrm{sun}}_{\nu_e\to\nu_e}(E)$
with neutrino energy $E$
is limited by 
$
\Delta P^{\mathrm{sun}}_{\nu_e\to\nu_e}(E)
\leq
\left( 1 - c_e \right)^2
$
\cite{BGG-AB-96,BGG-review-98}.
If $c_e$ is large as in the LS or LL regions of Fig.~\ref{4dis1}
or in the L region of Fig.~\ref{4dis2},
we have
$
P^{\mathrm{sun}}_{\nu_e\to\nu_e}
\geq
\left( 1 - a^0_e \right)^2 / 2
\simeq
1/2
$
and
$
\Delta P^{\mathrm{sun}}_{\nu_e\to\nu_e}(E)
\leq
(a^0_e)^2
\lesssim
10^{-3}
$,
for
$\Delta m^2_{41}=\Delta{m}^2_{\mathrm{LSND}}$
in the LSND range (\ref{LSND-range}).
Therefore,
$P^{\mathrm{sun}}_{\nu_e\to\nu_e}$
is bigger than 1/2 and
practically
does not depend on neutrino energy.
Since this is incompatible with the results of solar neutrino experiments
interpreted in terms of neutrino oscillations,
we conclude that the regions
LS and LL in Fig.~\ref{4dis1}
and the region L in Fig.~\ref{4dis2}
are disfavored by solar neutrino data,
as illustrated qualitatively by the vertical exclusion lines in
Figs. \ref{4dis1} and \ref{4dis2}.

Let us consider now the results of atmospheric neutrino experiments.
Small values of $c_\mu$ are incompatible with atmospheric neutrino oscillations
because in this case $\nu_\mu$ has
small mixing with
the two massive neutrinos responsible for atmospheric neutrino oscillations.
Indeed, the survival probability of
atmospheric $\nu_\mu$'s
is bounded by
$
P^{\mathrm{atm}}_{\nu_\mu\to\nu_\mu} \geq \left( 1 - c_\mu \right)^2
$
\cite{BGG-AB-96,BGG-review-98},
and it can be shown
that the Super-Kamiokande
up--down asymmetry of high-energy
$\mu$-like events generated by atmospheric neutrinos,
$
\mathcal{A}_\mu
=
0.311 \pm 0.043 \pm 0.01
$
\cite{Scholberg-99},
and
the exclusion curve of the Bugey $\bar\nu_e$
disappearance experiment
imply the upper bound
$ c_\mu \gtrsim 0.45 \equiv b^{\mathrm{SK}}_\mu $.
This limit is depicted qualitatively by the horizontal
exclusion lines in Figs. \ref{4dis1} and \ref{4dis2}.
Therefore,
we conclude that the regions
SS and LS in Fig.~\ref{4dis1}
and the small-$c_\mu$ parts
of the regions S and L in Fig.~\ref{4dis2}
are disfavored by atmospheric neutrino data.

Finally, let us consider the results of the LSND experiment.
In the schemes of class 2 the amplitude of
short-baseline
$
\stackrel{\makebox[0pt][l]
{$\hskip-3pt\scriptscriptstyle(-)$}}{\nu_{\mu}}
\to\stackrel{\makebox[0pt][l]
{$\hskip-3pt\scriptscriptstyle(-)$}}{\nu_{e}}
$
oscillations is given by
$ \displaystyle
A_{{\mu}e}
=
\bigg|
\sum_{k=1,2} U_{ek} U_{\mu k}^*
\bigg|^2
=
\bigg|
\sum_{k=3,4} U_{ek} U_{\mu k}^*
\bigg|^2
$
($A_{{\mu}e}$
is equivalent to
$\sin^2 2\vartheta$,
where $\vartheta$
is the two-generation mixing angle used in the analysis of
the data of short-baseline
$
\stackrel{\makebox[0pt][l]
{$\hskip-3pt\scriptscriptstyle(-)$}}{\nu_{\mu}}
\to\stackrel{\makebox[0pt][l]
{$\hskip-3pt\scriptscriptstyle(-)$}}{\nu_{e}}
$
experiments).
The second equality is due to the unitarity of the mixing matrix.
Using the Cauchy--Schwarz inequality we obtain
\begin{equation}
c_e \, c_\mu
\geq
A^\mathrm{min}_{{\mu}e} / 4
\qquad
\mbox{and}
\qquad
\left( 1 - c_e \right) \left( 1 - c_\mu \right)
\geq
A^\mathrm{min}_{{\mu}e} / 4
\,,
\label{Amuel-bounds}
\end{equation}
where
$A^\mathrm{min}_{{\mu}e}$
is the minimum value of the oscillation amplitude
$A_{{\mu}e}$
observed in the LSND experiment.
The bounds (\ref{Amuel-bounds})
are illustrated qualitatively in Figs. \ref{4dis1} and \ref{4dis2}.
One can see that
the results of the LSND experiment confirm the exclusion of
the regions
SS and LL in Fig.~\ref{4dis1}
and the exclusion
of the small-$c_\mu$ part of region S
and
of the large-$c_\mu$ part of region L
in Fig.~\ref{4dis2}.

Summarizing,
if
$ \Delta{m}^2_{41} \gtrsim 0.3 \, \mathrm{eV}^2 $
only the region SL in Fig.~\ref{4dis1},
with
\begin{equation}
c_e \leq a^0_e
\qquad \mbox{and} \qquad
c_\mu \geq 1 - a^0_\mu
\,,
\label{c-bounds-1}
\end{equation}
is compatible with the results of all neutrino oscillation experiments.
If
$ \Delta{m}^2_{41} \lesssim 0.3 \, \mathrm{eV}^2 $
only the large-$c_\mu$ part of region S in Fig.~\ref{4dis2},
with
\begin{equation}
c_e \leq a^0_e
\qquad \mbox{and} \qquad
c_\mu \geq b^{\mathrm{SK}}_\mu
\,,
\label{c-bounds-2}
\end{equation}
is compatible with the results of all neutrino oscillation experiments.
Therefore,
in any case $c_e$ is small and $c_\mu$ is large.
However,
it is important to notice that,
as shown clearly in Figs. \ref{4dis1} and \ref{4dis2},
the inequalities
(\ref{Amuel-bounds})
following from the LSND observation of
short-baseline
$
\stackrel{\makebox[0pt][l]
{$\hskip-3pt\scriptscriptstyle(-)$}}{\nu_{\mu}}
\to\stackrel{\makebox[0pt][l]
{$\hskip-3pt\scriptscriptstyle(-)$}}{\nu_{e}}
$
oscillations
imply that $c_e$ and $1-c_\mu$, albeit small,
have the lower bounds
\begin{equation}
c_e \gtrsim A^\mathrm{min}_{{\mu}e} / 4
\qquad \mbox{and} \qquad
1 - c_\mu \gtrsim A^\mathrm{min}_{{\mu}e} / 4
\,.
\label{c-bounds-3}
\end{equation}

\section{Long-baseline experiments}
\label{Long-baseline experiments}

The smallness of $c_e$ in the schemes A and B implies that
electron neutrinos do not oscillate
in atmospheric and long-baseline neutrino oscillation experiments.

\begin{figure}[t]
\begin{center}
\includegraphics[bb=85 550 443 765,width=0.8\linewidth]{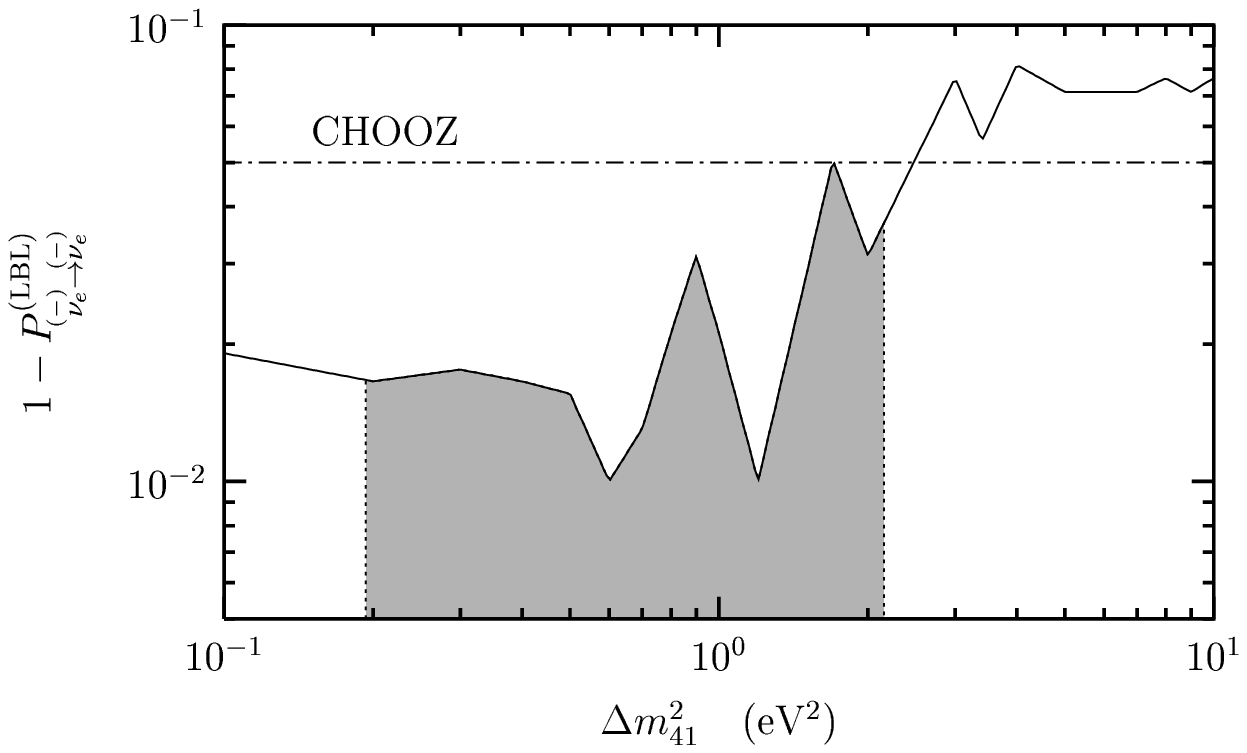}
\refstepcounter{figure}
\label{pee}
\\
\small
Figure \ref{pee}
\end{center}
\end{figure}

The transition probabilities
of electron neutrinos and antineutrinos
into other states in long-baseline experiments (LBL) are bounded by
\cite{BGG-bounds-98}
\begin{equation}
1 -
P^{(\mathrm{LBL})}_{\stackrel{\makebox[0pt][l]
{$\hskip-3pt\scriptscriptstyle(-)$}}{\nu_{e}}
\to\stackrel{\makebox[0pt][l]
{$\hskip-3pt\scriptscriptstyle(-)$}}{\nu_{e}}}
\leq
a^{0}_{e}
\left( 2 - a^{0}_{e} \right)
\,.
\label{05}
\end{equation}
The solid line in Fig.~\ref{pee} shows the corresponding
limit obtained from the 90\% CL exclusion plot of the Bugey experiment.
The shadowed region in Fig.~\ref{pee}
is allowed if
$\Delta{m}^2_{41}$
lies in the LSND range (\ref{LSND-range}).
The dash-dotted line in Fig.~\ref{pee} shows the
upper bound for the transition
probability of $\bar\nu_e$'s into other states
obtained from the final 90\%
exclusion plot of the CHOOZ \cite{CHOOZ} experiment
for $\Delta{m}^2_{\mathrm{atm}} \gtrsim 3 \times 10^{-3} \, \mathrm{eV}^2$
(the final 95\%
exclusion plot of the CHOOZ experiment gives
$
P^{(\mathrm{LBL})}_{\stackrel{\makebox[0pt][l]
{$\hskip-3pt\scriptscriptstyle(-)$}}{\nu_{e}}
\to\stackrel{\makebox[0pt][l]
{$\hskip-3pt\scriptscriptstyle(-)$}}{\nu_{e}}}
\lesssim 0.6
$).
One can see that the results of the CHOOZ experiment
agree with the upper bound (\ref{05}),
that is more stringent than the CHOOZ bound
for $\Delta{m}^2_{41}$ in the LSND range.

\begin{figure}[t]
\begin{center}
\includegraphics[bb=85 550 443 765,width=0.8\linewidth]{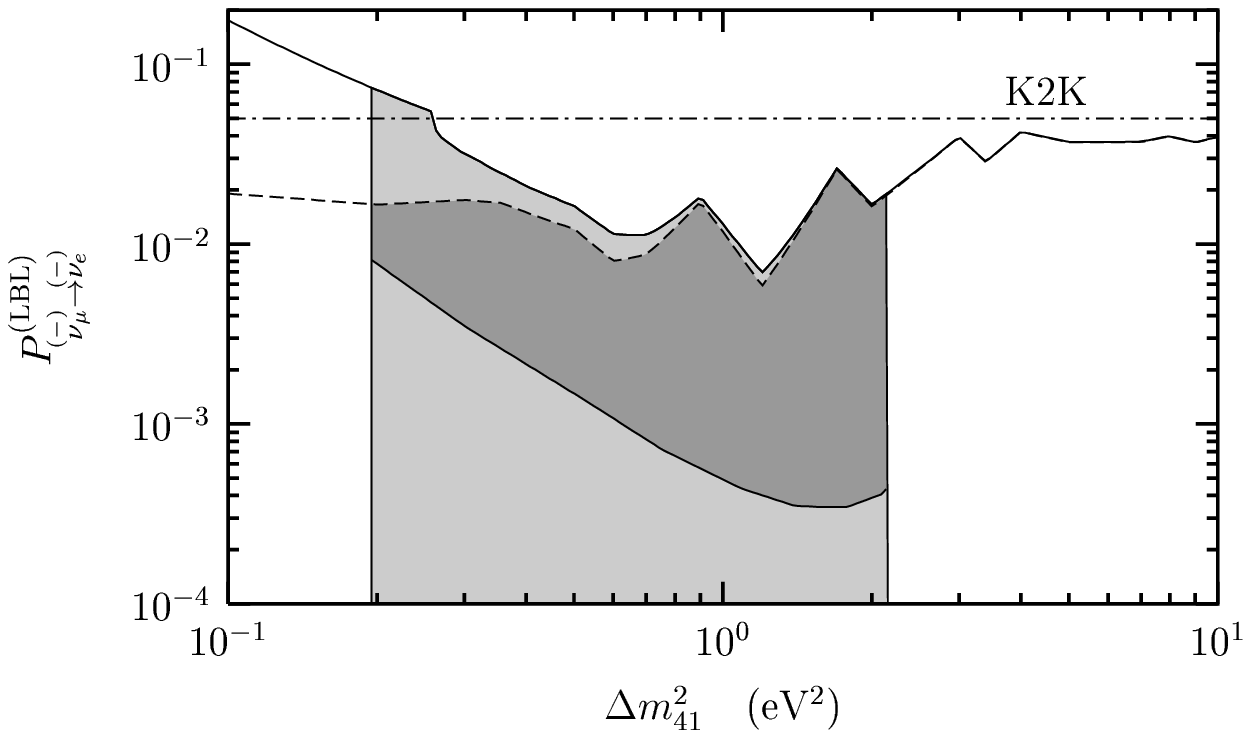}
\refstepcounter{figure}
\label{pk2k}
\\
\small
Figure \ref{pk2k}
\end{center}
\end{figure}

The probability of
$
\stackrel{\makebox[0pt][l]
{$\hskip-3pt\scriptscriptstyle(-)$}}{\nu_{\mu}}
\to\stackrel{\makebox[0pt][l]
{$\hskip-3pt\scriptscriptstyle(-)$}}{\nu_{e}}
$
transitions in vacuum in LBL experiments
is limited by
\cite{BGG-bounds-98}
\begin{equation}
\frac{1}{4} \, A^\mathrm{min}_{{\mu}e}
\leq
P^{(\mathrm{LBL})}_{\stackrel{\makebox[0pt][l]
{$\hskip-3pt\scriptscriptstyle(-)$}}{\nu_{\mu}}
\to\stackrel{\makebox[0pt][l]
{$\hskip-3pt\scriptscriptstyle(-)$}}{\nu_{e}}}
\leq
\min\!\left[
a^{0}_{e}
\left( 2 - a^{0}_{e} \right)
\, , \,
a^{0}_{e}
+
\frac{1}{4}
\,
A^{0}_{{\mu}e}
\right]
,
\label{06}
\end{equation}
where
$A^{0}_{{\mu}e}$
is the upper bound for the amplitude
$A_{{\mu}e}$
of short-baseline
$
\stackrel{\makebox[0pt][l]
{$\hskip-3pt\scriptscriptstyle(-)$}}{\nu_{\mu}}
\to\stackrel{\makebox[0pt][l]
{$\hskip-3pt\scriptscriptstyle(-)$}}{\nu_{e}}
$
transitions measured in accelerator neutrino experiments
and
$A^\mathrm{min}_{{\mu}e}$
is the minimum value of $A_{{\mu}e}$
observed in the LSND experiment.
The bound obtained with Eq.~(\ref{06})
from the 90\% CL exclusion plots of the Bugey
experiment
and
of the
BNL E776 \cite{BNL-E776} and KARMEN \cite{karmen-moriond-99}
experiments
is depicted by the dashed line
in Fig.~\ref{pk2k}.
The dark shadowed region is allowed by the results of the
LSND experiment,
taking into account the lower bound in Eq.~(\ref{06}).
The solid line in Fig.~\ref{pk2k} shows the upper bound on
$
P^{(\mathrm{LBL})}_{\stackrel{\makebox[0pt][l]
{$\hskip-3pt\scriptscriptstyle(-)$}}{\nu_{\mu}}
\to\stackrel{\makebox[0pt][l]
{$\hskip-3pt\scriptscriptstyle(-)$}}{\nu_{e}}}
$
in the K2K experiment
\cite{K2K}
taking into account matter effects
\cite{BGG-bounds-98}.
In this case there is no lower bound
and
the dark plus light shadowed regions are allowed
by the results of the LSND experiment.
The expected 90\% CL sensitivity
of the K2K long-baseline accelerator neutrino experiment
for
$\Delta{m}^2_{\mathrm{atm}} \gtrsim 3 \times 10^{-3} \, \mathrm{eV}^2$
is indicated in Fig.~\ref{pk2k}
by the dash-dotted line.
It can be seen that the results of short-baseline experiments
indicate an upper bound for
$
P^{(\mathrm{LBL})}_{\stackrel{\makebox[0pt][l]
{$\hskip-3pt\scriptscriptstyle(-)$}}{\nu_{\mu}}
\to\stackrel{\makebox[0pt][l]
{$\hskip-3pt\scriptscriptstyle(-)$}}{\nu_{e}}}
$
smaller than the expected sensitivity of
the K2K experiment,
unless
$\Delta{m}^2_{41} \simeq 0.2 - 0.3 \, \mathrm{eV}^2$.

Let us emphasize that the upper bounds
for the oscillation probabilities in long-baseline experiments
presented in
Figs. \ref{pee} and \ref{pk2k}
depend on $\Delta{m}^2_{41}$,
that is the mass-squared difference relevant for
oscillations in short-baseline experiment.
The transition probabilities measured in each long-baseline experiment
can be much smaller that the maximal one,
that lies below the upper bounds
in
Figs. \ref{pee} and \ref{pk2k},
if $\Delta{m}^2_{\mathrm{atm}}$
is much smaller than the
mass-squared difference to which the experiment is
most sensitive.

A further consequence of the smallness of $c_e$ and $1-c_\mu$
in the schemes A and B
is the existence of a stringent upper bound
for the size of CP or T violation that could be measured
in long-baseline experiments
in the $\nu_\mu\leftrightarrows\nu_e$
and $\bar\nu_\mu\leftrightarrows\bar\nu_e$ channels
\cite{BGG-CP-98}.
On the other hand,
the effects of CP violation
in long-baseline
$\nu_\mu\leftrightarrows\nu_\tau$
and
$\bar\nu_\mu\leftrightarrows\bar\nu_\tau$
transitions
can be as large as allowed by the unitarity of the mixing matrix
\cite{BGG-CP-98}.

\section{Conclusions}
\label{Conclusions}

We have seen that only the two four-neutrino schemes A and B
of class 2 in Fig.~\ref{4spectra}
are compatible with the results of all neutrino oscillation experiments.
These two schemes are equivalent for
the phenomenology of neutrino oscillations.
We have shown that the quantities $c_e$ and $1-c_\mu$
in the schemes A and B
are small.
Physically $c_\alpha$,
defined in Eq.~(\ref{04}),
quantify the mixing of the flavor neutrino $\nu_\alpha$
with the two massive neutrinos whose $\Delta{m}^2$
is relevant for the oscillations of atmospheric neutrinos
($\nu_1$, $\nu_2$ in scheme A
and
$\nu_3$, $\nu_4$ in scheme B).
Considering long-baseline neutrino oscillation experiments,
the smallness of of $c_e$ implies
stringent upper bounds
for the probability of
$
\stackrel{\makebox[0pt][l]
{$\hskip-3pt\scriptscriptstyle(-)$}}{\nu_{e}}
$
transitions into other states,
for the probability of
$
\stackrel{\makebox[0pt][l]
{$\hskip-3pt\scriptscriptstyle(-)$}}{\nu_{\mu}}
\to\stackrel{\makebox[0pt][l]
{$\hskip-3pt\scriptscriptstyle(-)$}}{\nu_{e}}
$
transitions
and for the size of CP or T violation effects in
$\nu_\mu\leftrightarrows\nu_e$
and
$\bar\nu_\mu\leftrightarrows\bar\nu_e$
transitions.


\begin{thebibliography}{10}

\bibitem{exp-sun}
B.T. Cleveland \textit{et al.} (Homestake Coll.), Astrophys. J. \textbf{496},
  505 (1998); K.S. Hirata \textit{et al.} (Kamiokande Coll.), Phys. Rev. Lett.
  \textbf{77}, 1683 (1996); W. Hampel \textit{et al.} (GALLEX Coll.), Phys.
  Lett. \textbf{B447}, 127 (1999); J.N. Abdurashitov \textit{et al.} (SAGE
  Coll.), astro-ph/9907113; Y. Fukuda (Super-Kamiokande Coll.), Phys. Rev.
  Lett. \textbf{82}, 2430 (1999).

\bibitem{exp-atm}
Y. Fukuda \textit{et al.}, Phys. Rev. Lett. \textbf{82}, 2644 (1999); Y. Fukuda
  \textit{et al.} (Kamiokande Coll.), Phys. Lett. \textbf{B335}, 237 (1994); R.
  Becker-Szendy \textit{et al.} (IMB Coll.), Nucl. Phys. B (Proc. Suppl.)
  \textbf{38}, 331 (1995); W.W.M. Allison \textit{et al.} (Soudan 2 Coll.),
  Phys. Lett. \textbf{B449}, 137 (1999); M. Ambrosio \textit{et al.} (MACRO
  Coll.), Phys. Lett. \textbf{B434}, 451 (1998).

\bibitem{LSND-Moriond-99}
G. Mills (LSND Coll.), Talk presented at the XXXIV$^{\mathrm{th}}$ Rencontres
  de Moriond \textit{Electroweak Interactions and Unified Theories}, Les Arcs,
  March 1999 
  (http://{\-}moriond.{\-}in2p3.{\-}fr/{\-}EW/{\-}transparencies).

\bibitem{BGG-review-98}
S.~M. Bilenky, C.~Giunti, and W.~Grimus,
\newblock (1998), hep-ph/9812360,
\newblock To be published in Progress in Particle and Nuclear Physics, Volume
  43.

\bibitem{Bugey-95}
Y.~Declais {\em et~al.},
\newblock Nucl. Phys. {\bf B434}, 503 (1995).

\bibitem{CHOOZ}
M. Apollonio \textit{et al.} (CHOOZ Coll.), hep-ex/9907037.

\bibitem{SK-atm}
Y. Fukuda \textit{et al.} (Super-Kamiokande Coll.), Phys. Rev. Lett.
  \textbf{81}, 1562 (1998); K. Scholberg (Super-Kamiokande Coll.),
  hep-ex/9905016.

\bibitem{BGG-AB-96}
S.M. Bilenky, C. Giunti and W. Grimus, Eur. Phys. J. \textbf{C1}, 247 (1998),
  hep-ph/9607372; Proc. of \textit{Neutrino '96}, Helsinki, June 1996, edited
  by K. Enqvist \textit{et al.}, p.~174, World Scientific, 1997,
  hep-ph/9609343.

\bibitem{Barger-variations-98}
V.~Barger, S.~Pakvasa, T.~J. Weiler, and K.~Whisnant,
\newblock Phys. Rev. {\bf D58}, 093016 (1998), hep-ph/9806328.

\bibitem{BGGS-AB-99}
S.~M. Bilenky, C.~Giunti, W.~Grimus, and T.~Schwetz,
\newblock Phys. Rev. {\bf D60}, 073007 (1999), hep-ph/9903454.

\bibitem{Giunti-70-99}
C. Giunti, hep-ph/9909465, Talk presented at \textit{Neutrino Mixing}, Meeting
  in Honour of Samoil Bilenky's 70$^{\mathrm{th}}$ Birthday, Torino, 25--27
  March 1999.

\bibitem{BGKP-96}
S.~M. Bilenky, C.~Giunti, C.~W. Kim, and S.~T. Petcov,
\newblock Phys. Rev. {\bf D54}, 4432 (1996), hep-ph/9604364.

\bibitem{CDHS-CCFR-84}
F. Dydak \textit{et al.} (CDHS Coll.), Phys. Lett. \textbf{B134}, 281 (1984);
  I.E. Stockdale \textit{et al.} (CCFR Coll.), Phys. Rev. Lett. \textbf{52},
  1384 (1984).

\bibitem{Scholberg-99}
K. Scholberg (Super-Kamiokande Coll.), hep-ex/9905016.

\bibitem{BGG-bounds-98}
S.~M. Bilenky, C.~Giunti, and W.~Grimus,
\newblock Phys. Rev. {\bf D57}, 1920 (1998), hep-ph/9710209.

\bibitem{BNL-E776}
L.~Borodovsky {\em et~al.},
\newblock Phys. Rev. Lett. {\bf 68}, 274 (1992).

\bibitem{karmen-moriond-99}
T.E. Jannakos (KARMEN Coll.), hep-ex/9908043.

\bibitem{K2K}
Y. Oyama (K2K Coll.), hep-ex/9803014; K2K WWW page:
  http://{\-}pnahp.{\-}kek.{\-}jp/.

\bibitem{BGG-CP-98}
S.~M. Bilenky, C.~Giunti, and W.~Grimus,
\newblock Phys. Rev. {\bf D58}, 033001 (1998), hep-ph/9712537.

\end{thebibliography}

\end{document}